\documentclass[10pt,journal,compsoc]{IEEEtran}

\usepackage{cite}
\usepackage{inputenc}
\usepackage{amsmath}
\usepackage{graphicx}
\graphicspath{{./}}
\usepackage{changepage}
\usepackage[colorinlistoftodos]{todonotes}
\usepackage{color,soul}
\usepackage{threeparttable}
\usepackage{setspace}
\usepackage{booktabs}
\usepackage{fancyhdr }
\usepackage{longtable}
\usepackage{caption}
\usepackage{blindtext}
\usepackage{scrextend}
\addtokomafont{labelinglabel}{\sffamily}
\usepackage{ragged2e}
\usepackage{enumitem}

\renewenvironment{description}[1][0pt]
  {\list{}{\labelwidth=0pt \leftmargin=#1
   }}
  {\endlist}

\begin{document}

\title{A Physical Unclonable Function with Redox-based Nanoionic Resistive Memory}
\author{Jeeson~Kim,~\IEEEmembership{Student Member,~IEEE,} 
Taimur~Ahmed, Hussein~Nili, Jiawei~Yang, Doo~Seok~Jeong,
Paul~Beckett,
Sharath~Sriram,~\IEEEmembership{Member,~IEEE,}
Damith~C.~Ranasinghe,~\IEEEmembership{Member,~IEEE,}
and~Omid~Kavehei,~\IEEEmembership{Senior Member,~IEEE}
\IEEEcompsocitemizethanks{\IEEEcompsocthanksitem J.~Kim, P.~Beckett and O.~Kavehei are with Nanoelectronic and Neuro-inspired Research Laboratory, RMIT~University, VIC~3000, Australia.\protect\\ E-mails: \{jeeson.kim,pbeckett,omid.kavehei\}@rmit.edu.au.
\IEEEcompsocthanksitem T.~Ahmed, S.~Sriram, J.~Kim and O.~Kavehei are with the Functional Materials and Microsystems Research Group, RMIT~University, VIC~3000, Australia.\protect\\
E-mails: \{taimur.ahmed,sharath.sriram\}@rmit.edu.au.
\IEEEcompsocthanksitem H.~Nili is with the Department of Electrical and Computer Engineering, University of California, Santa Barbara, CA~93106~USA.\protect\\ E-mail: hnili@ece.ucsb.edu.
\IEEEcompsocthanksitem J.~Yang is with Wenzhou Medical University, China.\protect\\ E-mail: yangjw@wibe.ac.cn.
\IEEEcompsocthanksitem D.~S.~Jeong is with the Electronic Materials Research Centre, Korea Institute of Science and Technology, 136-791 Seoul, Republic of Korea.\protect\\ E-mail: dsjeong@kist.re.kr.
\IEEEcompsocthanksitem D.~C.~Ranasinghe and J.~Kim are with the Auto-ID Labs, School of Computer
Science, The University of Adelaide, SA~5005, Australia.\protect\\ E-mail: damith.ranasinghe@adelaide.edu.au.
}
\thanks{Manuscript received ; revised . This work was supported by Australian Research Council under grant DP140103448 and National Natural Science Foundation of China under Grant 61501332.}
}

\markboth{}%
{Author \MakeLowercase{\textit{et al.}}: Journal title}

\IEEEtitleabstractindextext{%
\begin{abstract}
A unique set of characteristics are packed in emerging nonvolatile reduction-oxidation (redox)-based resistive switching memories (ReRAMs) such as their underlying stochastic switching processes alongside their intrinsic highly nonlinear current-voltage characteristic, which in addition to known nano-fabrication process variation make them a promising candidate for the next generation of low-cost, low-power, tiny and secure Physically Unclonable Functions (PUFs). This paper takes advantage of this otherwise disadvantageous ReRAM feature using a combination of novel architectural and peripheral circuitry. We present a physical one-way function, {\it nonlinear resistive Physical Unclonable Function} ({\it nr}PUF), potentially applicable in variety of cyber-physical security applications given its performance characteristics. We experimentally verified performance of Valency Change Mechanism (VCM)-based ReRAM in nano-fabricated crossbar arrays across multiple dies and runs. In addition to a massive pool of Challenge-Response Pairs (CRPs), using a combination of experimental and simulation, our proposed PUF shows a reliability of $98.67$\%, a uniqueness of $49.85$\%, a diffuseness of $49.86$\%, a uniformity of $47.28$\%, and a bit-aliasing of $47.48$\%.
\end{abstract}

}

\maketitle

\IEEEdisplaynontitleabstractindextext

\IEEEpeerreviewmaketitle

\section{Introduction}\label{sec:introduction}
\IEEEPARstart{R}{edox} based resistive memories (ReRAMs) are an emerging class of two-terminal nonvolatile memory technology. They are one of the most promising devices for conventional and unconventional information processing and memory applications~\cite{choi2015data, legenstein2015computer, prezioso2015training}. They can be integrated on-chip with conventional CMOS technology. Thanks to their highly nonlinear resistance behavior, their read-out speed can be programmed to be fast or slow, with higher or lower power consumptions, respectively. 
 As there are countless choices of material, these devices can also be fabricated with an aim for ultra-high density digital memories or behave like an analog memory with multiple stable states. Unlike SRAM, DRAM and FLASH technologies, these devices do not rely on charge storage and retention on a capacitor, rather they exploit a new type of underlying physics that is based on a mixed ionic-electronic conduction mechanism and hence, it is a resistance retention-based technology. These memories have also surpassed nonvolatile FLASH technology in almost any aspect; They offer orders of magnitude higher endurances and write speed than FLASH, they have eliminated FLASH memory\rq{}s need for high voltage supply and cumbersome erase procedure, and they offer all these advantages at a potentially lower volume fabrication cost~\cite{waser2007nanoionics, wong2015memory, coteus2011technologies, valov2013nanobatteries, waser2009redox, strukov2016endurance}. ReRAMs could be made potentially denser than FLASH and like FLASH they can go vertical (3D) and ensure zero-power consumption when on stand-by~\cite{tsunoda2007low,chen2015exploiting,hu2016physically}.

Security and privacy applications of such device as part of broader cyber-physical system industry ranging from nation-wide power grids, small scale health care system to Internet of Things (IoT) and solutions based on radio frequency identification (RFID)~\cite{konstantinou2015cyber,suh2007physical}. For instance, IoT demands challenging security solutions with features such as interoperability, scalability and lightweight process~\cite{suh2007physical,damiani2005new}. In such area, power and performance demanding environment, Physical Unclonable Functions (PUFs) could be a suitable solution~\cite{rajendran2015nano,vskoric2005robust, van2013anti,guajardo2008brand,ranasinghe2006confronting,devadas2008design,cole2008networked}.

A 19$^{\rm th}$ century scientist, Auguste Kerckhoffs once stated, {\it a system should be secure even if everything about the system, except the key, is a public knowledge}~\cite{auguste1883cryptographie}. Building on this principle, PUF\rq{}s very fundamental feature is the ability to remove the requirement of storing secrets and leave any other system feature as public knowledge~\cite{zhang2014exploiting,zhang2015puf}. Secrets in PUF are intrinsically encrypted in physical implementation randomness --e.g. silicon fabrication process, for which there is no way to reverse engineer the system but to characterize every single component of the system. It is therefore very difficult, if not impossible, to make an identical copy of the system even by the use of identical processes, facilities and material~\cite{zhang2015highly}. The basic idea of a PUF is {\it to gain advantage of otherwise disadvantageous physical system manufacturing non-idealities}. These non-idealities can be classified into {\it spatial} and {\it temporal} variations. Spatial variations include process variations such as dimensions, random dopant fluctuations, line-edge roughness, which manifest themselves as conventionally undesirable features such as threshold voltage variation and offset in CMOS and other electrical characteristics of solid-state devices. Temporal variation includes noise, supply power, temperature fluctuations, transient effects as common effects in both conventional and emerging technologies. One unique property that is unique to ReRAM devices, is their random oxygen vacancy profile. 

\begin{figure}[!t]
\centering
{\includegraphics[width=2.7in]{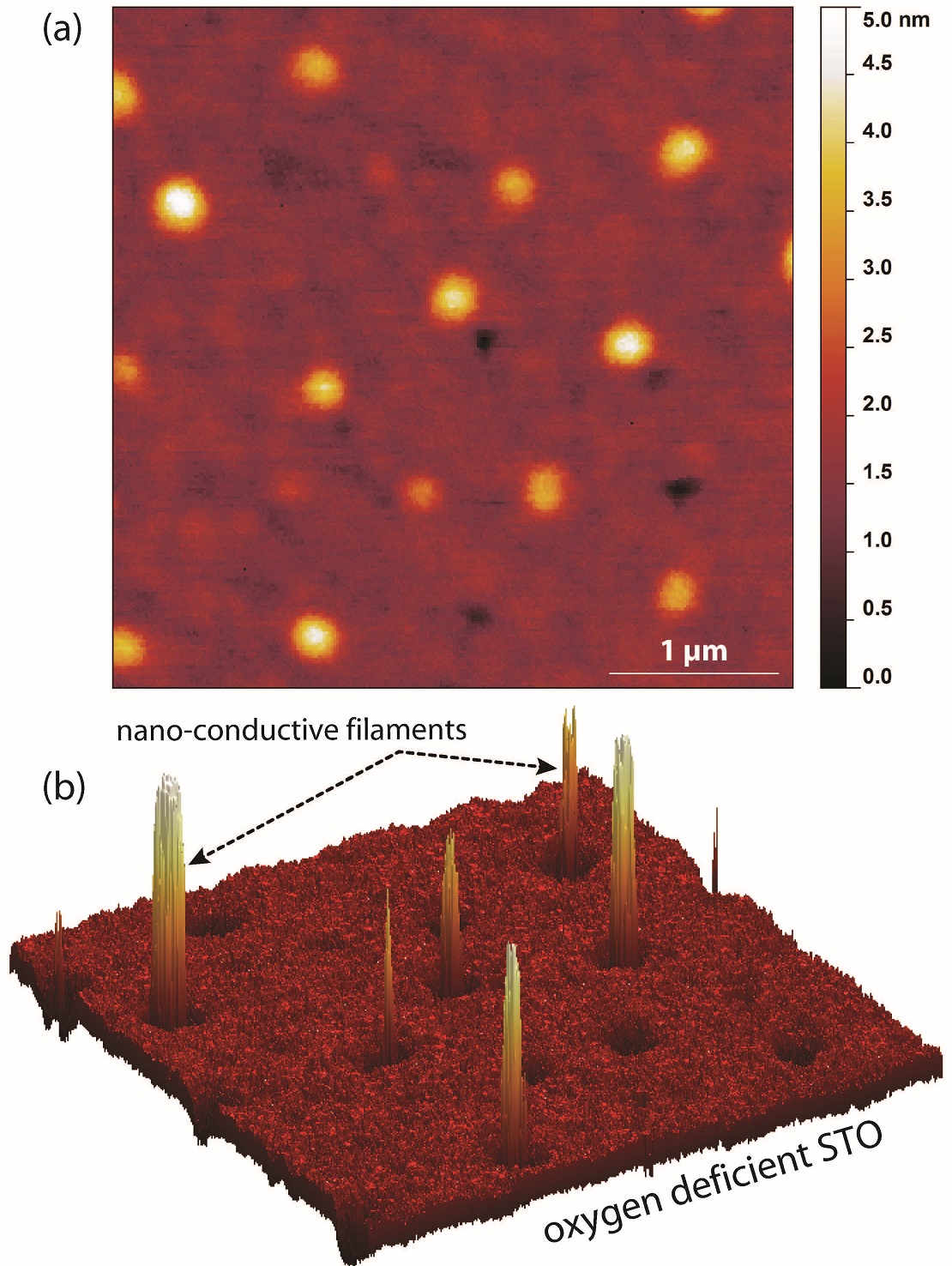}}
\caption{\label{fig:SPMb}In-situ Scanning Probe Microscopy (SPM) of conductivity is shown in 2D (a) and 3D (b) maps. It shows a top view of oxygen deficient amorphous SrTiO$_{3{\rm-}x}$ ($a$-STO) layer, also known as the switching layer, in our ReRAM devices after removing top metal electrode layers for this experiment. The pattern illustrates profile of nano-filaments of heights of 5 nm, which are conducting channels between two metal electrodes in ReRAMs. Formation and rupture of these nano-conductive filaments are possible when enough energy is delivered via an application of current. There are countless evidence, including our device characteristics, that there are two or more distinctive thresholds after which formation and rupture of these filaments take place and each time that occurs, the spatial pattern (location) of these filaments and also their height, shape and intensity could also be different. This adds a temporal feature to the oxygen vacancy profile. However, it is observed that when working significantly below --voltage-- thresholds, location, shape, height and intensity of these filaments are reliably fixed and unique to each device.}
\end{figure}

Fig.~\ref{fig:SPMb}(a) and (b), shows a direct evidence of oxygen vacancy profile in our devices (true for all ReRAMs)~\cite{nili2015donor}. This pattern is likely to change with every switching as the formation and rupture of nano-filaments in certain location is an stochastic phenomenon. This pattern is also varying from device-to-device. The tricky part about this spatio-temporal random oxygen vacancy profile in ReRAMs is that once programming is finished, the profile stays fixed under no or small magnitudes of energy delivered to our Valency Change Mechanism (VCM)-based ReRAMs. Temporal and transient aspects of this nano-conductive filament pattern is more profound when the device is switching or is in its Low Resistance State (LRS). It can also be concluded that when a filament becomes the main path of conducting current between electrodes, ReRAM\rq{}s LRS\rq{}s conductance is almost independent of device contact sizes and thier variation, while in High Resistance State (HRS), the nano-conductive filament pattern is fixed, unique to each device, and their height are much less than LRS\rq{}s nano-filaments. Dimensional and line-edge roughness variations are also mainly effecting HRS. Therefore, oxygen vacancy profile could be consider as a perfectly spatial parameter when a ReRAM device is in its HRS. It is important to note that discussions around oxide-trap-induced effects such as burst or random telegraph noise are outside the scope of this paper due to the relatively low frequency nature of the phenomenon.

These systematic, random and spatial variations in CMOS has been harnessed to produce tiny differences in identical circuit and system performance. For instance, small differences in {\it delay} has been the source of spatial randomness in arbiter PUFs (Arb-PUFs) and ring-oscillator PUFs (RO-PUFs)~\cite{lim2005extracting, gassend2004identification,suh2007physical}. SRAM-PUF is another example. SRAM\rq{}s {\it metastability} is widely used as another source of randomness. Process variation in implementation of SRAM\rq{}s latch and read-out/addressing transistors could make its switch to \lq\lq{}1\rq\rq{} or \lq\lq{}0\rq\rq{} more likely than the other after setting up a metasptable condition~\cite{guajardo2007fpga, holcomb2007initial, kumar2008butterfly, su2008digital}. While some CMOS PUFs are custom designed, many have been reported on field array logics such as Field-Programmable Gate Array (FPGA). That unfortunately created systematic bias issues as there is no ultimate control over interconnect length~\cite{ruhrmair2013puf}.

\begin{figure*}[htp]
\centering
\includegraphics[width=7.0in]{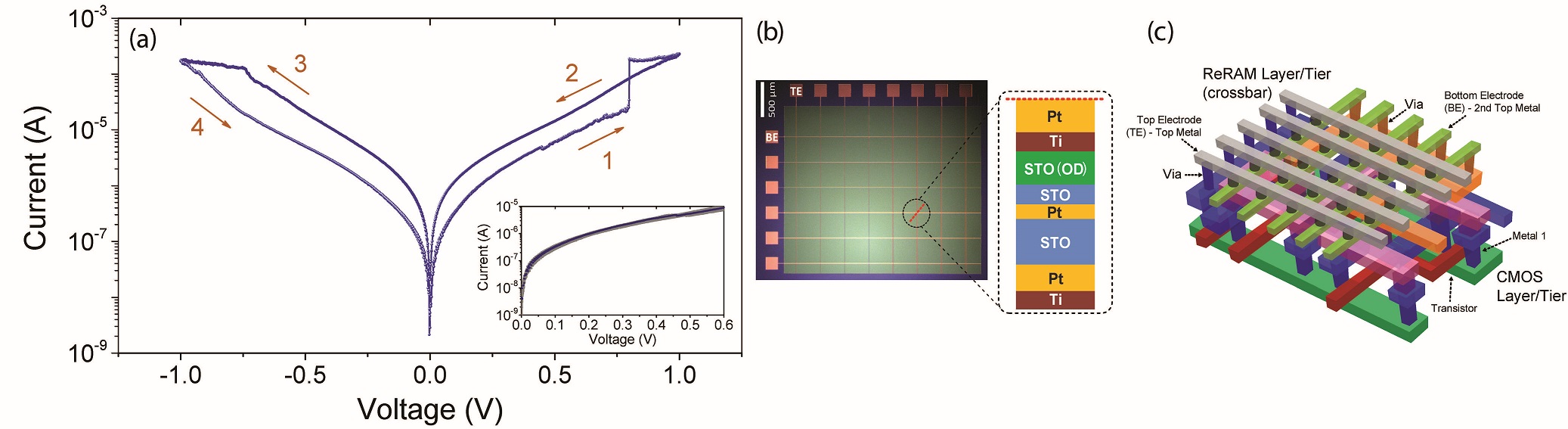}
\caption{\label{fig:fab1}ReRAM electrical characteristic and structure. (a) Experimental current-voltage ($I$-$V$) signature of our $a$-STO ReRAM bipolar switching behavior. As a representative signature curve of thousands of measured $I$-$V$ characteristics on multiple devices, this is measured by a DC double voltage sweeping. Arrows and numbers on the response $I$ curve indicate voltage sweeping directions and the order of sweeping operations. Inset: highlights nonlinearity of $I$-$V$ curve when device is in its HRS and applied voltages are significantly below threshold. These voltages are called READ voltages in this paper. (b) Optical microscope image of a $6\times 8$ ReRAM crossbar array and schematic of our ReRAM material stack. (c) Schematic of how ReRAM crossbar could be integrated with CMOS.}
\end{figure*}

While silicon technology has started a new ground-breaking wave of security primitive solutions, the game is still on for true random key generators that are highly secure, cheap, small and energy-efficient~\cite{darpaS}. The device-to-device randomly variant oxygen vacancy profile in VCM ReRAMs provides another dimension that if utilized appropriately could pave the way for implementation of highly secure nonvolatile memory (NVM)-based PUFs. In this work, we propose a novel PUF architecture based on nonvolatile ReRAM crossbar arrays. Contributions of this work include:
\begin{itemize}
\item Introducing a nonlinear ReRAM-based PUF and presenting its full analysis.
\begin{itemize}
\item This includes, assessment of important PUF metrics based on a mix of extensive experimental analysis and simulation.
\end{itemize}
\item Employing the idea of dummy cells and arrays to strengthen {\it nr}PUF against side-channel power monitoring attacks.
\end{itemize}

The paper is organized as follows: Section~\ref{nrPUF} discuss properties of ReRAMs and proposes {\it nr}PUF and its circuit and architectural level operation. {\it nr}PUF characterization and experimental results of fabricated ReRAM arrays are illustrated. Section~\ref{Eval} analyses {\it nr}PUF performance metrics and discusses simulation results. Finally, Section~\ref{conclusion} summarizes the work.

\section{Nonlinear Resistive PUF}\label{nrPUF}

\subsection{Electrical properties of ReRAM}\label{STO}

Measured signature bipolar switching behavior of our VCM ReRAM devices is depicted in Fig.~\ref{fig:fab1}(a) at room temperature. Device switching characteristic becomes available when an irreversible electro-forming step is completed. An electro-forming step forces the device to switch from its pristine state to its LRS. Beyond that point the device is capable of switching between its LRS and HRS, when enough energy is delivered to the device in form of applied current. Our device SET (HRS$\rightarrow$LRS) and RESET (LRS$\rightarrow$HRS) switching thresholds are around 800 mV and -750 mV, respectively, as it is shown in Fig.~\ref{fig:fab1}(a). For electro-forming a maximum sweep voltage range of $2.5$ to $3.2$~V and current compliance range $100$ to $500$~$\mu$A was used. The switching behavior is known to be caused by the formation and rupture of one or more filamentary paths through the oxide layer between Top Electrode (TE) and Bottom Electrode (BE)~\cite{niliNano, nili2015donor}. This is shown in Fig.~\ref{fig:SPMb}. Switching sequence (1-4) is shown in Fig.~\ref{fig:fab1}(a) and the device was initially in its HRS. Electrical characterization and measurement data was gathered with Keithley 4200 Semiconductor Characterization System. It is worths noting that electro-forming and its impact on spatio-temporal characteristics of oxygen vacancy profile is an interesting topic which is outside the scope of this paper and underpins further investigation.

Using standard photolithography we designed and fabricated a stack of the following materials to implement our VCM ReRAM devices. A $20$~nm Pt and its $5$~nm Ti adhesion layer are deposited on a SiO$_2$/Si substrate as BE using electron-beam evaporation. An amorphous SrTiO$_{3}$, $a$-STO, ($33$~nm) film is subsequently sputtered through a shadow mask and in the next step, a $5$~nm Pt as buffer metal layer is e-beam evaporated on the $a$-STO layer. Then, two layers of $a$-STO films were sputtered in different conditions; A $30$~nm oxygen deficient (OD) $a$-STO layer on a normal $3$~nm $a$-STO. Finally, a Pt/Ti ($20$~nm/$10$~nm) is formed by e-beam evaporation as TE. All deposition steps were processed at room temperature and a crossbar optical image and its material stack is shown in Fig.~\ref{fig:fab1}(b). The crossbar array consists of $8$ columns of TEs and $6$ rows of BEs. When a voltage below switching threshold (Fig.~\ref{fig:fab1}(a)\rq{}s inset), known as READ voltage, is applied to the TE, it produces a current that can be read-out from BE. Full details on fabrication process can be found in Refs.~\cite{nili2015donor,niliNano}. ReRAM switching layer in our devices is an amorphous OD SrTiO$_{3{\rm-}x}$ ($a$-STO), where $x$ represents the level of oxygen deficiency created by a combination of processes within the material stack during fabrication and engineered by a detailed micro/nano-fabrication development recipe. 

\begin{figure}[!t]
\centering
\includegraphics[width=3.4in]{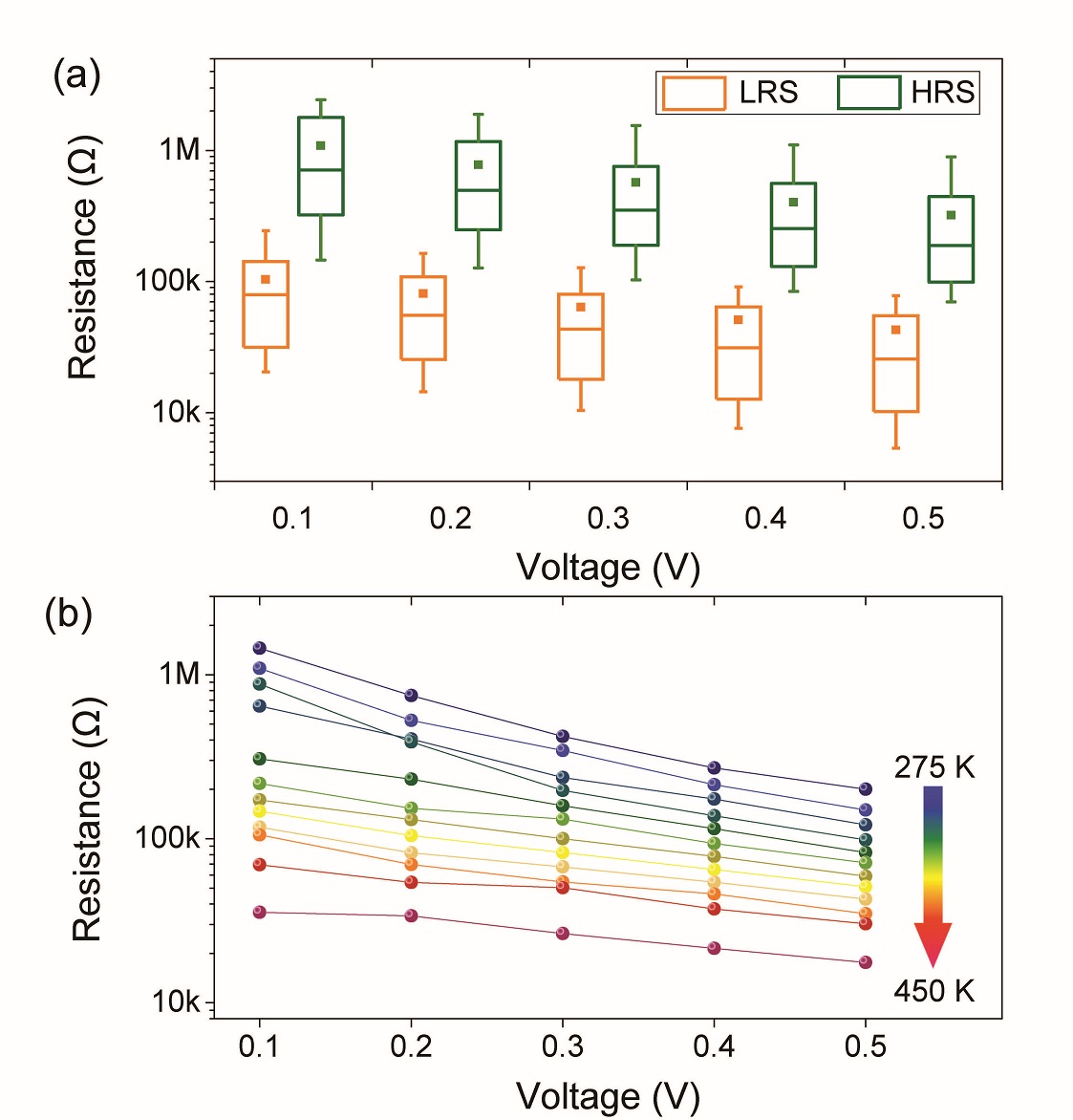}
\caption{\label{fig:fab}(a) Device-to-device (D2D) variation in HRS and LRS. State resistance variation of HRS and LRS are extracted from $58$ devices at different READ voltages between $0.1$ and $0.5$~V. As Fig.~\ref{fig:fab1}(a)\rq{}s inset suggests, nonlinearity of $I$-$V$ characteristics causes semi-exponential increase in HRS current with every $100$~mV increase in voltage. Therefore, as READ voltage increases, R$_{\rm OFF}$/R$_{\rm ON}$ ratio decreases. (b) Resistance systematic variation induced by temperature change from, near zero degree of Celsius, 275$^{\circ}$~K, to 450$^{\circ}$~K.}
\end{figure}

To highlight Fig.~\ref{fig:fab}(a) shows one main device-to-device (D2D) variation of our VCM ReRAM HRS and LRS at different READ voltages. In this paper, we only use HRS. Measurements remark that HRS is widely distributed over a decade in the range of $100$~k$\Omega$ to $1$~M$\Omega$. Fig.~\ref{fig:fab}(b) demonstrates temperature dependence of the ReRAM cell. It shows thermal activation of current transport through the cells in the temperature range 275$^{\circ}$ to 450$^{\circ}$~K. Although measured resistance transition over multiple devices is substantial, the behavior suggests a trend that can be considered in our peripheral read-out circuitry for {\it nr}PUF.

\subsection{Circuit and Architecture}

\begin{figure*}[htp]
\centering
{\includegraphics[width=6.8in]{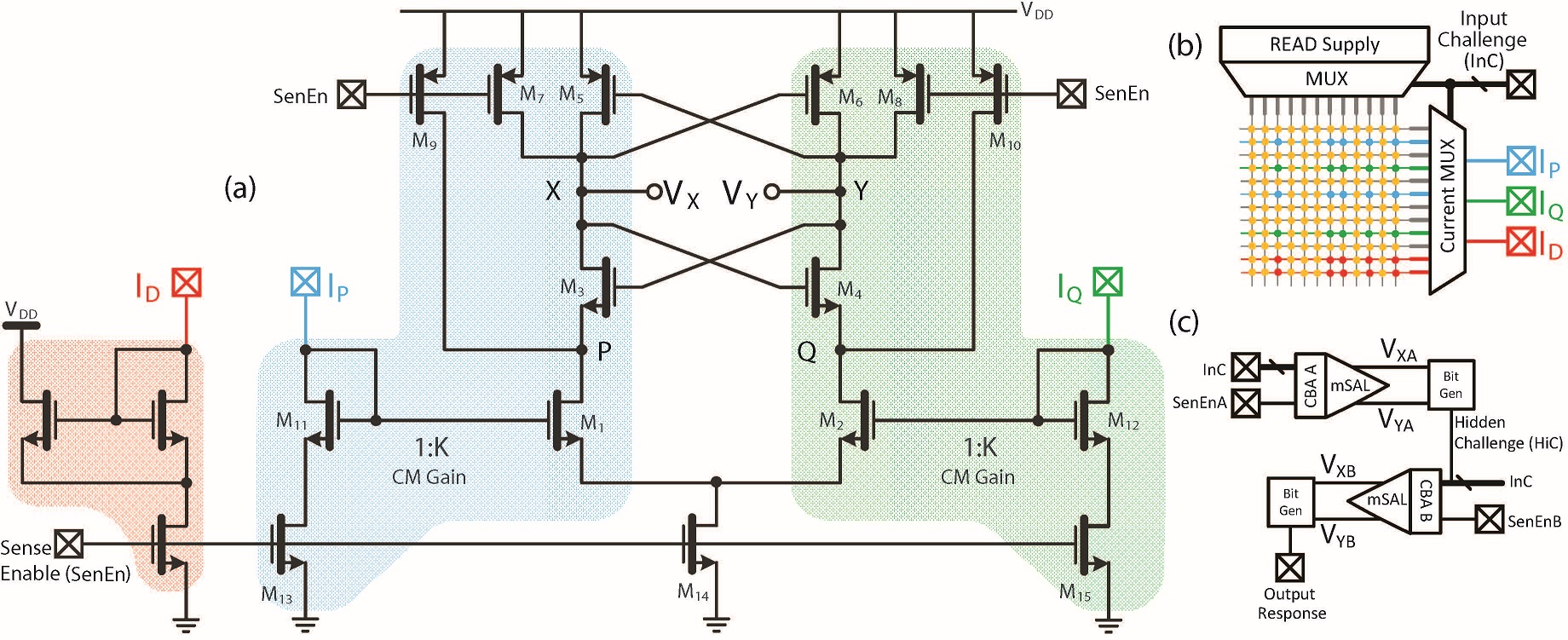}}
\caption{\label{fig:arch}Proposed {\it nr}PUF block diagram, interconnections and read-out circuitry. (a) Represents a modified StrongARM Latch (mSAL) that its original design and analysis could be found in Refs.~\cite{razavi2015strongarm,kobayashi1992current}. Transistors M$_{\rm 7-10,13-15}$ are acting as pre-charge and sensing mode triggers. This sensing enable (SenEn) signal could identify throughput of the system. The aim is to have an ideal probability of 50\% for each response bit to be either 1 or 0 at each phase that SenEn=0. We want this randomness to be dominated by VCM ReRAM devices and not the mSAL. The current mirrors for $I_{\rm P}$ and $I_{\rm Q}$ that are drawn from a crossbar array (CBA), (b), are assumed to have current mirror (CM) gain of $K$=1. The highlighted red part, $I_{\rm D}$, represents a dummy section that aims to confuse power signal and reduce its SNR in order to further diminish {\it nr}PUF protection against side-channel power monitoring attacks. (b) Illustrates a CBA with its relevant analog/current and digital multiplexers (Current MUX, aMUX, and MUX) accepting {\it q}-bit input challenge (InC) as input and selects a subset of columns (out of $M$) to be connected to READ voltage supply and a subset of rows (out of $N$) to be connected to $I_{\rm P}$, $I_{\rm Q}$ and $I_{\rm D}$. All ReRAMs cells are programmed to be in their HRS. (c) shows {\it nr}PUF clock diagram with two CBAs and mSALs. Part A\rq{}s output, whether one or {\it l}-bit is considered as hidden challenge as it directly influences, alongside InC, response bit generation at the output of Part B.}
\end{figure*}

The proposed {\it nr}PUF structure is shown in Fig.~\ref{fig:arch}. The overall system architecture consists of multiple VCM ReRAM crossbars, two sense amplifiers and bit generators. The system accepts parallel streams of 64-bit inputs that each is called a challenge and produces 64-bit output that is called a response. Each challenge produces 1-bit response. Fig.~\ref{fig:arch}(a) illustrates a modified StrongARM (mSAL) latch. We added transistors M$_{\rm 13}$, M$_{\rm 14}$ and M$_{\rm 15}$ to better control the flow of current when sensing is enabled using the signal, SenEn. The current mirror and control transistor in the left hand side and highlighted in red are serving dummy cells/arrays and are not part of the mSAL circuit and we discuss its effects on supply power signal-to-noise ratio (SNR) later in the paper. Original idea and full description of the sensing circuitry can be found in Refs.~\cite{razavi2015strongarm,kobayashi1992current}. The mSAL circuit consists of two identical parts, highlighted in green and blue that are competing to own the output, $V_{\rm x}$ and $V_{\rm y}$. Assuming negligible mismatch between peer transistors (e.g. M$_{\rm 1}$ and M$_{\rm 2}$), the state of the latch should be identified by the mismatch between $I_{\rm P}$ and $I_{\rm Q}$. As Fig.~\ref{fig:arch}(b) illustrates, these two currents are directly passing through a selection of multiple ReRAM devices which are all programmed in their HRS. Due to the randomly different oxygen vacancy profile of these devices, one of the currents will be higher than the other, which means  voltages at nodes P and Q will not be identical. That causes an unbalance in the current that is drawn by M$_{\rm 1}$ and M$_{\rm 2}$ after a pre-charge mode (SenEn=0) that charged nodes P, Q, X and Y to $V_{\rm DD}$. The unbalance will push the latch (transistors M$_{\rm 3-6}$) towards $V_{\rm x}$=$V_{\rm DD}$, $V_{\rm y}$=0 or $V_{\rm x}$=0, $V_{\rm y}$=$V_{\rm DD}$.

One of the most important systematic bias that needs to be mitigated is the offset generated as the result of mismatch between current mirror pairs, M$_{\rm 1,11}$ and M$_{\rm 2,12}$, and M$_{\rm 3-6}$ of the latch. It is known that in conventional StrongARM circuit, dominant contributors to the offset are M$_{\rm 1}$ and M$_{\rm 2}$~\cite{razavi2015strongarm}. In this case, we need to extend that set to include M$_{\rm 11}$ and M$_{\rm 12}$. Due to the fact that our architecture uses only  two mSAL circuits, as shown in Fig.~\ref{fig:arch}(c), there are plenty of room for mitigating M$_{\rm 1,2}$ offset contribution. It is very well-known that such offset in a FET (field effective transistor) is the direct result of mismatch in threshold voltages which is the consequence of process variation. According to the renowned Pelgrom\rq{}s Law, 
\begin{equation}
\sigma_{\Delta V_T}\propto \frac{1}{\sqrt{WL}},
\end{equation}
where $W$, $L$ and $\sigma_{\Delta V_T}$ representing length and width of transistor channel and standard deviation of threshold voltage mismatch, sampled from thousands of pairs~\cite{kinget2005device, pelgrom1989matching}, to avoid creating a systematic bias in our CMOS, M$_{\rm 1,11,2,12}$ should be as large as possible.

In the proposed PUF, input challenge either directly or through a set of linear-feedback shift registers (LFSRs), which are not shown in Fig.~\ref{fig:arch}(b) diagram, randomly activates 1, 2, 3, 4 or 5 columns out of $N$ columns and exactly two rows out of $M$ rows for each of the $I_{\rm P,Q,D}$ currents. In this paper, $N$=$M$=$128$, therefore, a massive pool of Challenge-Response-Pairs (CRPs) are expected as it is shown in Section~\ref{Eval}. Other array characteristics such as sneak-current paths (array parasitic currents) are data and addressing pattern dependent, and hence, are included in our analysis but their individual role were not studied in this paper. Fig.~\ref{fig:arch}(c) demonstrates that the input challenge (InC) is applied to two crossbar arrays A and B (CBA A and CBA B) and mSALs produce relevant output bits. For every operation, output bit of the first part, called Hidden Challenge (HiC) in this paper, influences selection of rows and columns in the second part. As its name suggests, it acts as a hidden challenge that participate in response bit generation and it could be multiple bits for different structures and requirements. While all parts are directly involved in response bit generation, the part highlighted in red, dummy, aim to confuse power consumption signal in order to reduce the adversarial chance in using side-channel power monitoring.

\subsection{Operation}
\begin{figure}[!t]
\centering
\includegraphics[width=3.1in]{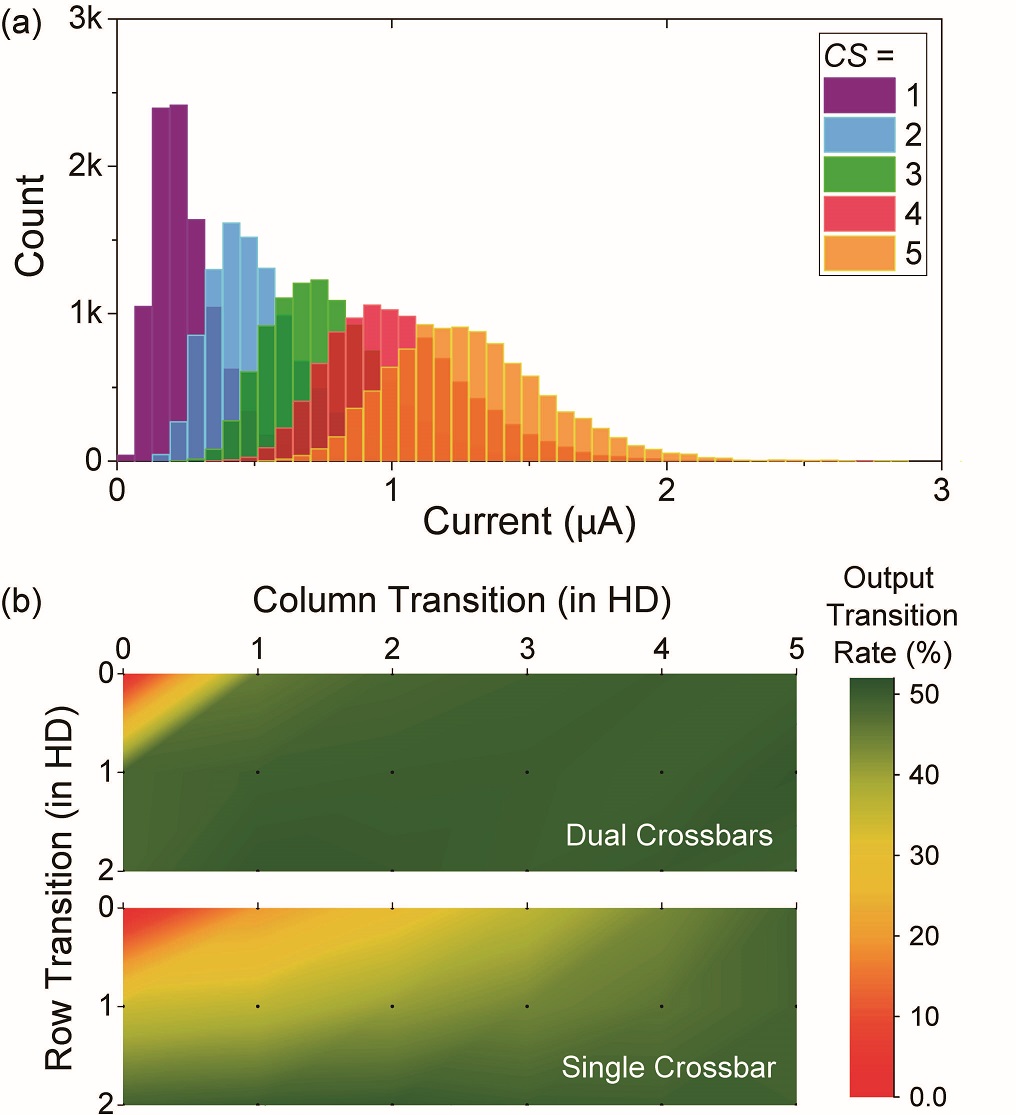}
\caption{\label{fig:Ava}Cells read-out distribution and avalanche behavior of {\it nr}PUF. (a) Row cells read-out current distribution for different $CS$. As $CS$ increases from 1 and 5, read-out current distribution becomes wider and its deviation ($\sigma$), therefore, increases from $132$~nA to $290$~nA. (b) Using the method we demonstrated in (b), this part shows a comparison between output transition rates when we have only one crossbar in the system (no HiC) compared to system shown in Fig.~\ref{fig:arch}(a) with dual crossbar arrays. The results are shown for $CS$=5 and x-axis demonstrates number of changes in selection of columns. When column transition value is $j$, it means our five selected columns, out of 128 available columns, are different from a reference selection of five columns with HD=$j$. For all column transitions, we select two rows ($l$=1 in Equ. (\ref{NCRP})). The y-axis demonstrates number of changes in selection of a pair of rows. When row transition value is $k$, it means our two selected rows, out of 128 available rows, are different from a reference selection of two rows with HD=$k$. The color-map shows output transition rate for each choice of row and column transitions compared to a reference of two row and five column selections. The more green the map is the closer the outcome is to the ideal case of a balance response, which means a 50\% output transition rate.}
\end{figure}

READ voltage of {\it nr}PUF operation is chosen from the set of READ voltages highlighted in Fig.~\ref{fig:fab}. As our read-out is current based, we aim to choose the lowest READ voltage possible to guarantee no destruction to the stored state. Due to inherent variation of ReRAM crossbar arrays, conductances of cells are widely distributed and that variation is ultimately translated to read-out current. Selected output of ReRAM crossbars are set as:

\begin{equation}
I_{{\rm row},i}=\sum_{k=1}^{CS}g_{i,j_k}V_{\rm READ}^{},
\end{equation}
where $g_{i,j_k}$ denotes conductance of cell located at ($i, j_k$) node, $i$ represents a device row location selected by analog/current multiplexer (aMUX), $j_k$ is a column location by decoder block and $CS$ is total number of columns selected. The aMUX utilizes a group of transmission gates for passing analog current inputs to its output. For temperature considerations, a temperature sensing circuitry could be beneficial alongside Fig.~\ref{fig:arch}(a). After read-out, current distribution is the root source of PUF uniqueness. In Section~\ref{STO}, we showed individual ReRAM cell resistance distribution, see Fig.~\ref{fig:fab}(a). In this work, we fixed $CS$=$5$. It utilizes a wider distribution than one or two column selection methods. Suppose $I_1, I_2, \ldots, I_{CS}$ are $CS$ number of independent random variables (a cell read-out current) with mean $\mu_1, \mu_2, \ldots, \mu_{CS}$ and variance $\sigma_1^2, \sigma_2^2, \ldots,\sigma_{CS}^2$. Then the mean and variance of the linear combination $I_{\rm row}=\sum_{k=1}^{CS}{I_k}$ are defined as: 
\begin{equation}
\mu_{I_{row}}=\sum_{k=1}^{CS}\mu_k
\end{equation}
and 
\begin{equation}
\sigma_{I_{row}}^2=\sum_{k=1}^{CS}\sigma_k^2,
\end{equation}
respectively. This shows that $I_{\rm row}$ distribution as well as its standard deviation increase with higher $CS$. This is shown in Fig.~\ref{fig:Ava}(a). Since {\it nr}PUF deals with comparison of electrical characteristics (linear sum of $CS$ number of cells' read-out current), the wider variation distribution provides the advantage of reducing possibility that selected comparator objects are in indistinguishable range. The high $CS$ method has another merit in that it increases the challenge space as well as prevents revealing of the PUF's variation fingerprint to adversaries attempting to characterize the PUF. Column and row selection on CBA A is entirely driven by a $q$-bit challenge using decoder and aMUXs block, see Fig.~\ref{fig:arch}(b) and (c). 

A CMOS unit selects $2\times l$ rows and each provides $I_{\rm A, row}$, and performs $l$ pairs of current comparison in order to deliver $l$-bit HiC as an input of CBA B\rq{}s CMOS unit. In this work $l$=1, and hence two rows in each operation is selected. While column selection is based on $q$-bit challenge ideally through decoders and LFSRs, an internally generated $l$-bit HiC participates in selection of CBA B\rq{}s columns and rows in order to produce, $I_{\rm B,row}$s, and finally generates $1$-bit final response for {\it nr}PUF. The number of selection can be adjusted considering the size of ReRAM crossbar array and can be set as $\log_2\binom{M}{2}$, where $M$ is the number of rows in CBA. Total number of CRPs ($N_{\rm CRP}$) also depends on the size of ReRAM crossbar arrays and it can be estimated as:

\begin{equation}
N_{\rm CRP}={\binom{N}{CS}} \times {\binom{M}{2}}\times l, \label{NCRP}
\end{equation} 
where $M,~N$ are the sizes of CBA ($M\times N$). It is worth recalling that $CS$ is the number of selected columns and $l$ is the HiC bit length.

\section{Performance evaluation}\label{Eval}

\subsection{Avalanche characteristic}\label{Avalanche}
A PUF should be a one-way function that ideally there is no link between its input and output. When this property is achieved, it becomes nearly impossible to guess challenge bits by observing corresponding response bits and vice versa~\cite{rueppel1985correlation}. CRPs should also be unrelated, so that knowing one CRP has no impact on predicting other unknown CRPs regardless of their similarity~\cite{maiti2012robust}. This can be better described as following;

\begin{enumerate}
\item Response bits should have similar probability over challenge space~\cite{rostami2014robust}. For each response bit, the probabilities of being \lq\lq{}1\rq\rq{} or \lq\lq{}0\rq\rq{} should likely to be equal, i.e, $Prob(r=1)=Prob(r=0)=50$\%. 
\item Response bit transition rate to a set of challenges with Hamming Distance (HD)~$=i$ should be $50$\%, where 1$\leq i \leq q$ and $q$ is a length of a challenge. Consider $q$-bit reference challenge ($C_{\rm Ref}$) and a challenge ($C_i$) with HD=1 from $C_{\rm Ref}$. When $i\in \{1,\ldots,S\}$ all with HD=1 from $C_{\rm Ref}$, we have a vector of size $S$ of responses to those challenges. To meet the strict avalanche criterion (SAC) requirement, $r_i\oplus r_{\rm Ref}$, which represents whether a transition in $i^{\rm th}$ response bit occured compared to the reference response ($r_{\rm Ref}$), should result in a balanced vector of $S$ responses. It means we should ideally observe a balanced (50\%) number of \lq\lq{}1\rq\rq{}s and \lq\lq{}0\rq\rq{}s in the output vector. In other words, $Prob(r_S\oplus r_{\rm Ref}$=$1)$=50\%~\cite{webster1985design}.
\end{enumerate}

Although it is very difficult, if not impossible, to prove {\it unclonability} mathematically, literature has shown that some PUFs are predictable~\cite{majzoobi2008testing,majzoobi2009techniques,ruhrmair2010modeling,ruhrmair2013puf,tajik2014physical}. For high immunity to these attacks, it is desirable to have the two properties mentioned above~\cite{lim2005extracting,suh2007physical}. In most PUFs, the first property may be obtained, but the second, avalanche behavior, is more difficult. This is particularly the case for linear Arb-PUF structure. The example utilizes sequence stages of four terminal switches with two inputs and two outputs. Each switch is controlled by a single bit which identifies the switch configuration. The fundamental idea is that, due to process variations, delays of an identical signal at the input arriving at two different output pins are slightly different. Every stage contains two multiplexers connecting inputs to outputs. In a low-throughput delay-based PUF architecture, independence among CRPs is hard to achieve. There exist attempts to design nonlinear PUF architectures and examples includes XOR PUF~\cite{suh2007physical} and Feed-Forward Arb-PUFs (FF~Arb-PUFs)~\cite{gassend2004identification,lim2005extracting}. XORing in Arb-PUFs is a powerful method to randomize this irregular output to generate a balanced output. An XOR PUF consists of multiple Arb-PUFs and an XOR function which XORs the responses of Arb-PUFs. They show improvement on SAC after adding the XOR function~\cite{majzoobi2012slender}. Another example, FF~Arb-PUFs, utilize one or few switch(es) that are independent of input. It means that feed-forward creates some hidden information and the PUF achieves a higher degree of complexity~\cite{gassend2004identification,majzoobi2009techniques}. 

In Fig.~\ref{fig:Ava}(c), a PUF with single crossbar array structure shows a biased output bit transition rate, and this is the particular case of low HD between row and column selections. Compared to single crossbar, the proposed {\it nr}PUF, dual crossbars, provides a significantly improved response bit stream balance  and SAC.

\subsection{Attacks}\label{sec:attacks}
Possible attacks on PUFs are various both on software and hardware level~\cite{roelthesis}. For PUFs with limited number of CRPs, $N_{\rm CRP}$, such as SRAM-PUFs, it may be possible to characterize the entire structure by direct probing and/or {\it side}-{\it channel} power monitoring attack~\cite{helfmeier2013cloning,ruhrmair2010strong}. It is also known that a high $N_{\rm CRP}$ makes {\it model-building attacks} possible more effective via machine learning analysis on large collection of output data. This immediately implies that PUFs with small throughputs will be less vulnerable to machine learning attacks. ReRAM-based PUFs may have the potential to mitigate these issues by providing adjustable throughput e.g. by adjusting sense amplifier, nanowire and overall parasitic capacitance at the output of each crossbar. Also, they adopt a random selection of a subset of memory cells and comparing the current passing through them in a total analog fashion~\cite{zhang2015optimizating,chen2015exploiting}. Using this method, a PUF with $M\times N$ crossbar size obtains at least as $N$ times as many challenges as RO-PUF with $M$ number of RO stages~\cite{zhang2015optimizating}.

\subsubsection{Simple and differential power analysis attacks} One of effective security threat targeting system implementation of cryptographic algorithms is side-channel attacks. Power analysis has been effectively used against different sensitive items such as smart-card microprocessors \cite{kocher2011introduction,messerges2002examining}. Differential Power Analysis (DPA) goal is to extract correlations between data and supply power fluctuations. Therefore, in systems that generating \lq\lq{}1\rq\rq{} and \lq\lq{}0\rq\rq{} in the output consume different current DPA will be an effective statistical tool that given enough traces is cable of extracting tiny correlations. Other power analysis techniques include correlation power analysis based on the Hamming distance model and partitioning power analysis \cite{kocher2011introduction,brier2004correlation,fei2012statistical,luo2011algorithmic,fei2014statistics}. 

The proposed {\it nr}PUF is using differential current analysis as the means to generate output bit therefore therefore correleation between the total current consumption and output bit diminishes. Additionally, we have exploited dummy arrays (as explained in Section~\ref{nrPUF}), in order to reduce any potential correlations even further. The relatively low-cost and effective performance of ReRAM technology offers the possibility of dummy array introduction to enable the capability of confusing power consumption pattern.

Power analysis attacks in general could be evaluated by signal-to-noise ratio (SNR) between the single-bit
unit power consumption and the standard deviation of power leakage \cite{fei2014statistics}. In {\it nr}PUF, we analyzed power consumption for generation of 2000 output bits. An input challenge dependent selection of 5 columns with 20 ReRAMs identifies the output bit, while one or more ReRAM on dummy arrays are randomly selected at the same time in order to achieve confusion. Due to random nature of selection of dummy arrays and devices involved in the process, invasive attacks such as laser cutting of interconnects would unlikely result in a functioning {\it nr}PUF without dummy devices. 

\begin{figure}[!t]
\centering
\includegraphics[width=3.2in]{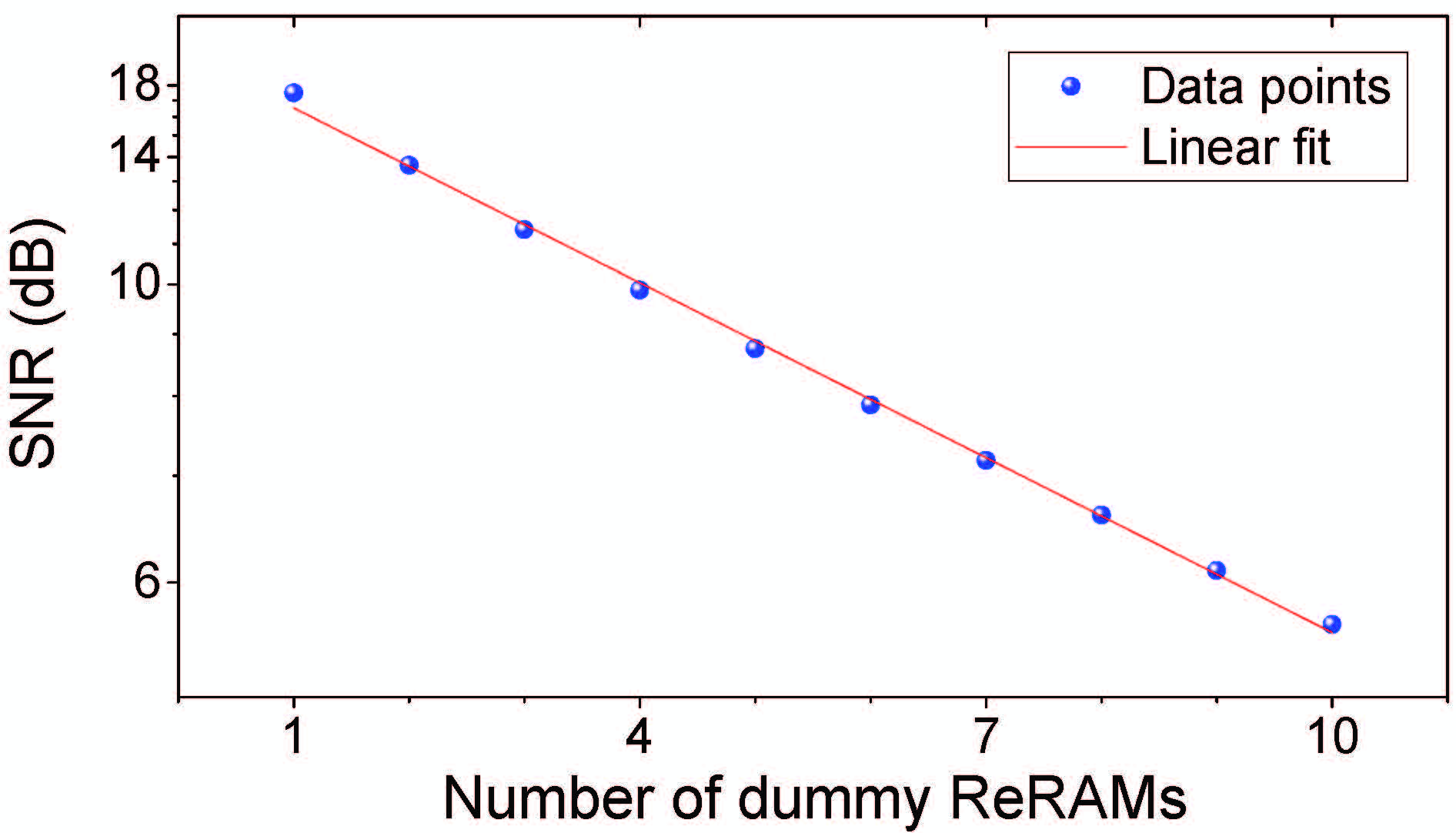}
\caption{\label{fig:Power}Role of dummy ReRAMs in lowering Signal-to-Noise Ratio (SNR) in an attempt to make supply power signal unrecoverable and unrelatable to output bit generation. Number of dummy ReRAMs is directly relevant to the magnitude of $I_{\rm D}$ in Fig.~\ref{fig:arch}(a) and (b).}
\end{figure}

Fig.~\ref{fig:Power} illustrates our SNR analysis result as a function of number of dummy devices involved in confusing power signal. As expected, the more the number of dummy devices are the lower SNR becomes and therefore it is possible to adjust such performance for different applications according to their sensitivity.

\subsection{PUF metrics}
We evaluated {\it nr}PUF against key PUF metrics in this part. Extensive circuit level Cadence simulations were followed up with rigorous Matlab analysis considering experimental data collected form a wide range of identical devices on same or different dies. Measured variations in current were fed into these simulations and devices were working under minimum READ voltage to be similar to experiments. There assumed noise and uncertainty on supply power line and existence of faulty devices (e.g. stuck-at-ON) in both crossbars. The following lists our considerations for analysis:

\begin{itemize}
\item {There is 10\% 3$\sigma$ READ supply voltage variation at any READ voltages,}
\item {A temperature fluctuation of $\pm$10$^{\circ}$~K at any working temperature,}
\item {An undetectable current difference of $\Delta I$=$\pm20$~nA, where $\Delta I$=$I_{\rm P}$-$I_{\rm Q}$,}
\item {90\% of HRS programmed devices were successful, therefore, 10\% of ReRAMs are assumed to be stuck-at-ON (in their LRS range, see Fig.~\ref{fig:fab}(a)), and}
\item {Measured ReRAM\rq{}s HRS variations have lognormal distribution, see Fig.~\ref{fig:fab}(a). These data were imported into analytical analysis flow to evaluate {\it nr}PUF.}
\end{itemize}
We use the following notations and definitions for {\it nr}PUF evaluation:

\begin{description}[0.7cm]
\item[$p$] {Number of PUF instances.}
\item[$n$] {Number of response bits.}
\item[$tr$] {Number of trials on the same PUF instance.}
\item[$r_{i, j}$] {$j^{\rm th}$ bit of $i^{\rm th}$ response.}
\item[$c$] {Number of challenges.}
\end{description}

\subsubsection{Hamming Weight (HW) test} 
\begin{figure}[!t]
\centering
\includegraphics[width=3.0in]{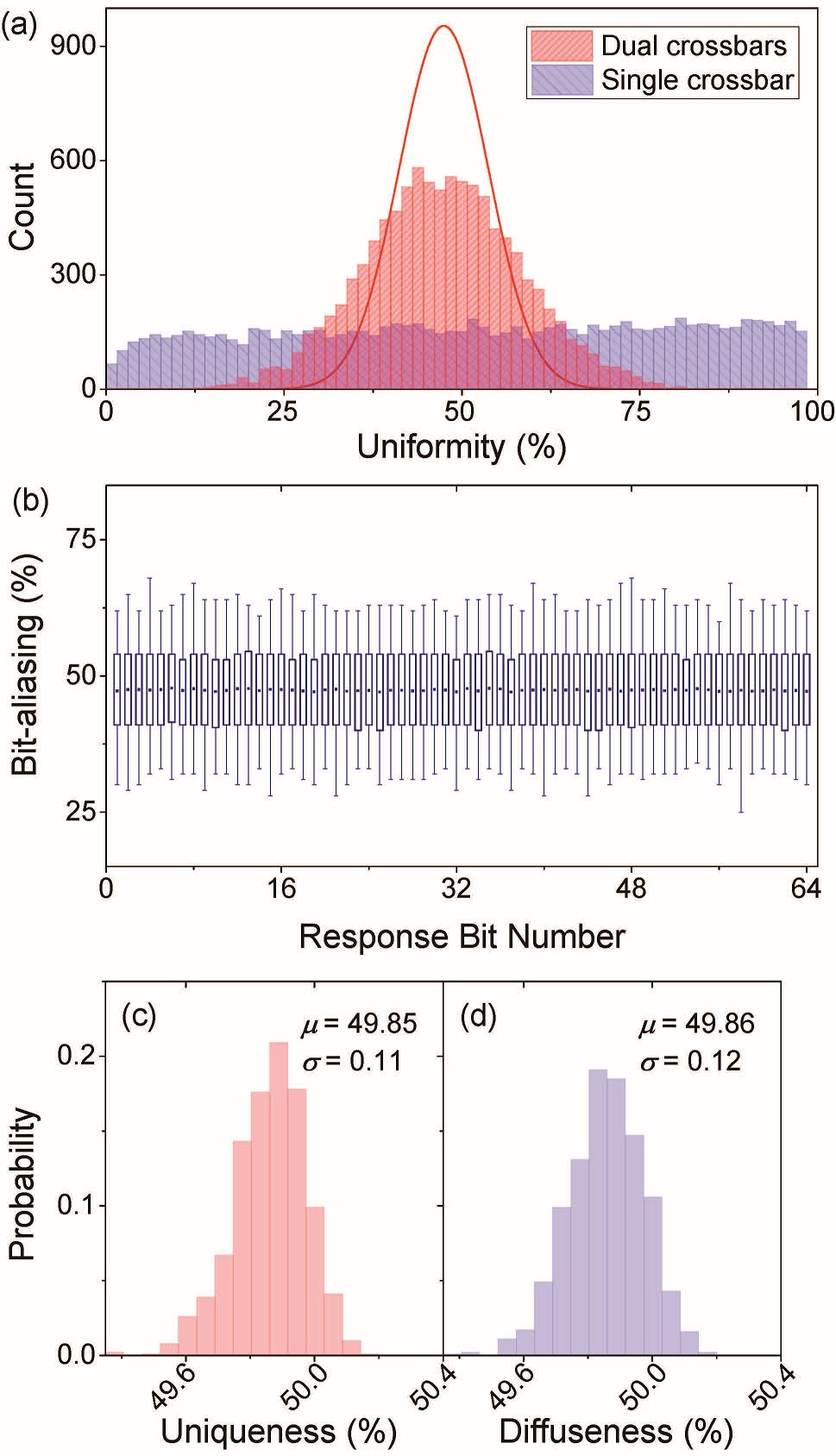}
\caption{\label{fig:eval}{\it nr}PUF Performance evaluations. (a) Worst-case uniformity (UF) comparison of response bit-stream of a {\it nr}PUF and a single crossbar structure. Red curve shows ideal distribution for UF, and it is clear that {\it nr}PUF is closer to the ideal UF than single crossbar-based PUFs. {\it nr}PUF\rq{}s UF shows a well-balanced \lq\lq{}0\rq\rq{}s and \lq\lq{}1\rq\rq{}s in its response bit-stream. (b) {\it nr}PUF\rq{}s bit-aliasing is shown. Each bit of {\it nr}PUF responses are assessed by calculating BA over 1000 PUF instances. The result indicate a well-balanced ratio of \lq\lq{}1\rq\rq{}s and \lq\lq{}0\rq\rq{}s across responses. (c) Uniqueness (UQ) and (d) diffuseness (DI) of {\it nr}PUF are demonstrated under a number of extreme temporal and transient nature uncertainty, such as supply voltage, temperature and sensing margin fluctuations.}
\end{figure}

HW test calculates inter- and intra-PUF responses in order to detect bit bias toward \lq\lq{}0\rq\rq{} or \lq\lq{}1\rq\rq{}. HW tests include uniformity (UF) and bit-aliasing (BA). Average UF and BA results are shown in Fig.~\ref{fig:eval}(a) and (b) and both are closely distributed near 50\%.

{\it Uniformity} (UF) is an intra-response HW assessment to evaluate a balance of \lq\lq{}0\rq\rq{}s or \lq\lq{}1\rq\rq{}s in a response vector. Ideally, UF should show a perfect balance. UF is defined as: 
\begin{equation}
{\rm UF}=\frac{1}{n}\sum_{j=1}^{n}r_{i, j}\times100\%,
\end{equation}
where $r_{i, j}$ is $j^{\rm th}$ bit of an $n$ bit response to $i^{\rm th}$ challenge. In Fig.~\ref{fig:eval}(a), red distribution curve represents the best-case UF of {\it nr}PUF when random challenges (zero or slightly larger than zero correlation between them) generate a response. It is closely distributed near to its ideal UF of 50\%. The worst-case UF is when challenges are not random and have high similarities with HD$_{challenge}\leq5$ considering a challenge length of 64-bit. Results shows the worst-case {\it nr}PUF, is normally distributed with $\mu$ of $47.28$\% and standard deviation of $11.09$\%. In contrast, UF of single crossbar structure is poorly centered and is rather uniformly distributed. This shows {\it nr}PUF could better satisfy desirable SAC behavior.

{\it Bit-Aliasing} (BA) is a measure that shows the degree of similarity across responses from different PUFs (inter-HW). Ideally, a PUF should avoid identical responses, hence, BA should be $50$\%. BA can be calculated as:
\begin{equation}
{\rm BA}=\frac{1}{p}\sum_{i=1}^{p}r_{i, j}\times100\%,
\end{equation}
where $r_{i, j}$ is $j^{\rm th}$ bit of an $n$ bit response from an $i^{\rm th}$ PUF instance. It is shown that average BA of {\it nr}PUF is $47.48$\% with deviation of $5.03$\%.

\subsubsection{Hamming Distance (HD) test} HD test calculates the HD of inter- and intra-PUF responses in order to assess how unique PUFs are. HD tests include uniqueness (UQ) and diffuseness (DF). Average UQ and DF results are shown in Fig.~\ref{fig:eval}(c) and (d) and both are closely distributed near 50\%. 

{\it Uniqueness} (UQ) is an inter-PUF HD test and an indicator of the PUF\rq{}s information bits that can be extracted by evaluating a degree of difference between responses of different PUFs to identical challenges. Truly random PUF should achieve UQ close to the ideal value of 50\%. Average UQ is defined as:
\begin{equation}
{\rm UQ}=\frac{1}{\binom{p}{2}}\sum_{i=1}^{p-1}\sum_{j=i+1}^{p}\frac{{\rm HD}(R_{i}, R_{j} )}{n}\times100\%,
\end{equation}
where HD($R_{i}, R_{j}$) is the HD between $n$ bit responses to a challenge from a pair of $i^{\rm th}$ and $j^{\rm th}$ PUF instances.

{\it Diffuseness} (DF) is an intra-PUF HD measurement, is to analyze a degree of response difference from different sets of challenges applied to the same PUF~\cite{diffus}. DF is defined as:
\begin{equation}
{\rm DF}=\frac{1}{\binom{c}{2}}\sum_{i=1}^{c-1}\sum_{j=i+1}^{c}\frac{{\rm HD}(R_{i},R_{j})}{n}\times100\%,
\end{equation}
where HD($R_{i}, R_{j}$) is the HD between $n$ bit responses to a pair of $i^{\rm th}$ and $j^{\rm th}$ challenge from a PUF instance.

\subsection{Reliability}\label{Reliability:sec}
PUFs are expected to demonstrate high {\it reliability}. Reliability shows PUF\rq{}s ability to reproduce same response to the same challenge over time and under significant spatio-temporal variations. In other words, it is defined as the probability that response bit $r_t$ that is generated at time $t$ to be reproduced at a $\Delta t$ later and $r_t=r_{t+\Delta t}$. An ideal PUF should provide 0\% difference in its responses to identical challenges and this is represented by Bit Error Rate (BER) definition below:
\begin{equation}
{{\rm BER}}=\frac{1}{\binom{tr}{2}}\sum_{i=1}^{tr-1}\sum_{j=i+1}^{tr}\frac{{\rm HD}(R_{i},R_{j} )}{n}\times100\%,
\end{equation}
where HD($R_{i}, R_{j}$) is the HD between responses to $i^{\rm th}$ and $j^{\rm th}$ application of a challenge to a PUF. Ideal reliability (RE) is $100$\% and is defined as:
\begin{equation}
{\rm RE}=100\%-{\rm BER}.
\end{equation}

\begin{figure}[!t]
\centering
\includegraphics[width=3.0in]{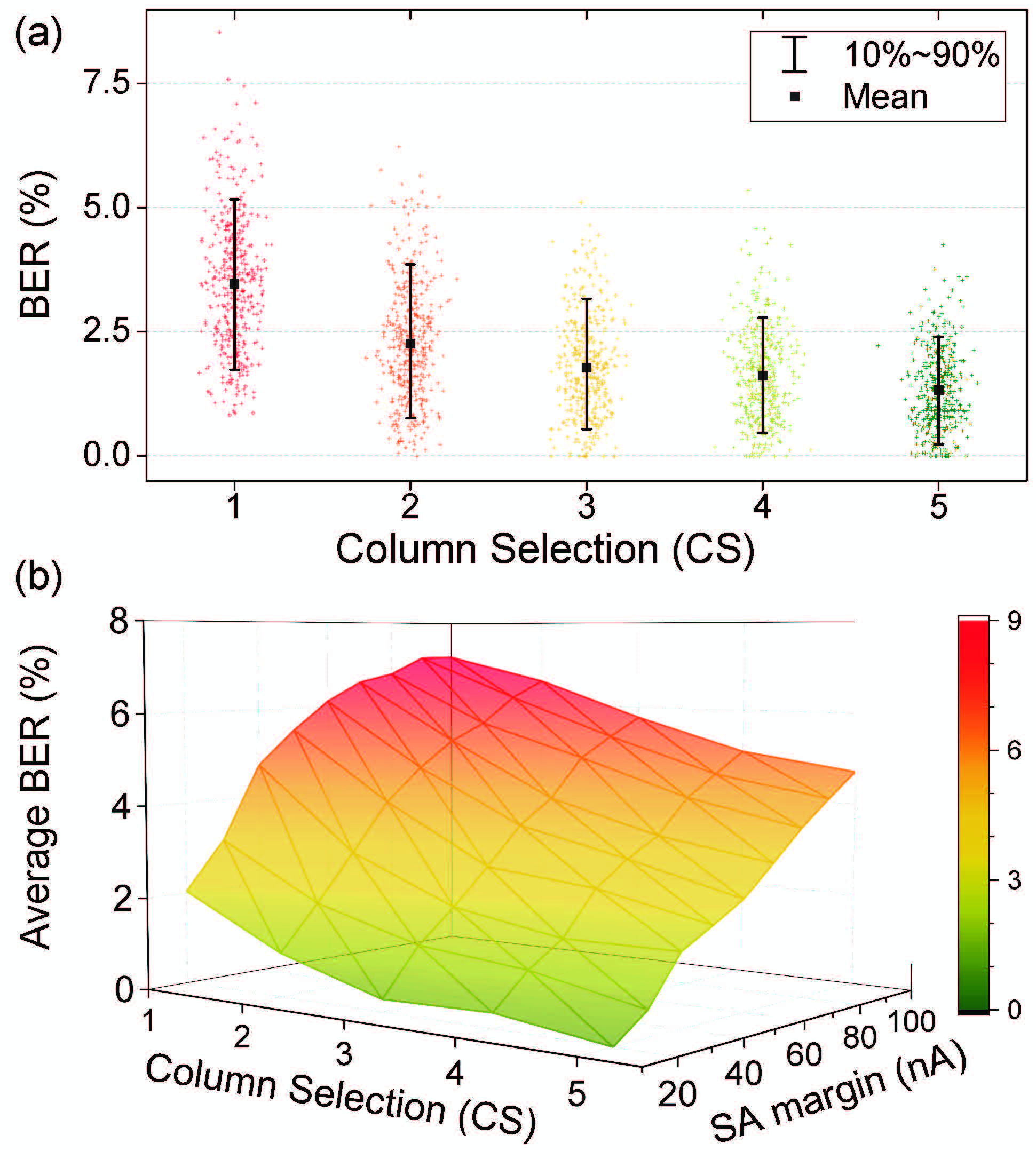}
\caption{\label{Reliability:fig}Bit Error Rate (BER) of {\it nr}PUF. (a) Reliability of {\it nr}PUF at a range of column number selection ($CS$) choices between 1 and 5 under supply voltage, temperature and sensing margin fluctuations. Reliability is significantly improved as $CS$ increases. (b) Average BER over multiple {\it nr}PUF analysis is shown as a function of $CS$ and mSAL error margin of $\Delta I$=$\pm10$ to $\pm100$~nA.}
\end{figure}

\begin{table*}[!t]
\centering
\begin{threeparttable}
\caption{\label{Comp}Comparison of crossbar PUFs.}
\small
\begin{tabular}{lcccccc}
\toprule
{} & NanoPPUF & M-PUF& MemristorPUF & {\it mr}SPUF & CPR-PUF  & {\it nr}PUF \\ \toprule
Reference&\cite{rajendran2012nano}&\cite{rose2013write} & \cite{koeberl2013memristor} &\cite{gao2015memristivecrypto, gao2015mrpuf} & \cite{chen2015exploiting} & This work \\ \midrule
Crossbar& 4$\times$4 & 8 bits & 1MB cells & 128$\times$128 & 1024$\times$1024& 2$\times$128$\times$128 \\ 
Minimum cell size & $6F^2$ & -- & $4F^2$ & $4F^2$ &  $4F^2$ & 2$\times$$4F^2$ \\ 
Memory state & LRS/HRS & LRS/HRS & -- & LRS & HRS & HRS \\ 
Uniqueness (\%) & $49$ & $49.85$ & \begin{tabular}[c]{@{}c@{}}$\sim48/50/55$ \\ (BC/Typ/WC)\end{tabular} & $50.07$ & $\sim49.95$ & $49.85$ \\ 
Reliability (\%)& -- & -- & -- & \begin{tabular}[c]{@{}c@{}}$92.5$\\ (WC)\end{tabular}& $\sim98$ & $ 98.67$ \\ 
Diffuseness (\%)& $49$ & -- & -- & $49.96$ & -- & $49.86$ \\ 
Uniformity (\%)& -- & $49.99$ & -- & $50.76$ & -- & $47.28$ \\ 
Bit-aliasing (\%)& -- & $49.99$ & -- & $49.99$ & -- & $47.48$ \\ [1ex]
CRPs calculation& -- & -- & -- & $\frac{N\times{\binom{M}{i}}\times{\binom{M-i}{i}}}{2}$ & ${\binom{N}{2}}\times{\binom{M}{2}}$ & ${\binom{N}{5}}\times{\binom{M}{2}} \times \log_2 {\binom{M}{2}}$ \\ [1ex]
Total CRPs$^\dag$& -- & -- & -- & $\sim3.7\times10^{18}$ & $\sim10^6$ & $\sim 2.7 \times 10^{13}$ \\ 
\bottomrule
\end{tabular}
\vspace*{0.1cm}\scriptsize{
\begin{tablenotes}
\item [] BC: Best-case 
\item [] $F$: ReRAM feature size
\item [] Typ: Typical-case 
\item [] WC: Worst-case 
\item [] {\dag}: Calculated for $N$=$M$=$128$ and $i$=$5$ for {\it mr}SPUF.
\item [] {--} Not mentioned
\end{tablenotes}}
\end{threeparttable}

\end{table*}

The reason for adopting a group of ReRAMs instead of a single device is to raise immunity against temporal variations. Although a single device comparison method has obvious advantages of consuming lower power, it has a poor reliability. When a PUF response stability is not guaranteed, the system requires an additional error correction module integrated with the PUF device and it increases costs, throughput and overall power consumption~\cite{devadas2013secure,oren2013effectiveness}.

Based on current distribution results, we evaluated reliability of {\it nr}PUF under different conditions. For each measurement set, we use $500$ random challenges and each challenge is repeated for $50$ trials in a PUF instance. Results clearly show the advantage of selecting multiple columns ($CS$=$5$) over one or two column(s) ($CS$=$1$ or $2$) selection method. Scattered cross symbols in Fig.~\ref{Reliability:fig}(a) represent average BER of $50$ trials to $500$ challenges. We assume SA margin for this work is $20$~nA. Mean value of BER $\mu_{\rm BER}$, which is $3.45$\% for $CS$=$1$ reduces by increasing $CS$. This is $2.39$\%, $1.87$\%, $1.61$\% and $\mu$ $1.33$\% for $CS$=2, 3, 4 and 5, respectively. As $CS$ and mSAL\rq{}s sensitivity increases, lower BER could be achieved (see Fig.~\ref{Reliability:fig}(b)). Table~\ref{Comp} presents a comparison between different proposed ReRAM PUFs. As the Table suggests, the proposed {\it nr}PUF could potentially achieve a closer performance metrics to the ideal, while all cited works have used similar mix of experimental-simulation analysis.

Excluding peripheral circuitry contribution, experimentally measured worst-case power consumption per ReRAM per response bit considering READ voltage of $100$~mV results power consumptions as low as $100$~nW. While simple estimation of power consumption based on this figure will be far from realistic total power consumption, specially by mSAL, it raises confidence in applicability of {\it nr}PUF. It is worth noting ReRAM arrays consume almost zero power while on stand-by. Data retention at the mentioned READ voltages has also been guaranteed for years at 85$^{\circ}$~C. Unlike start-up issues with SRAM-PUF~\cite{cortez2012modeling}, we believe {\it nr}PUF should provide a more reliable power-up phase thanks to their non-volatility and long data retention. According to our experimental observations of fastest pulse measurements using Keithley 4225-PMU, peripheral circuitry would dominate {\it nr}PUF throughput, which can be designed to have a range of operational speeds including slow read-outs as suggested in Ref.~\cite{ruhrmair2011applications}.

\section{Conclusion}\label{conclusion}
In summary, we present a novel {\it nr}PUF based on measured data collected from a range of ReRAM devices on one or multiple dies, fabricated under identical conditions. {\it nr}PUF utilizes a relatively simple ReRAM crossbar structure, minimizing its design phase to nanofabrication masks design. To improve unpredictability, {\it nr}PUF utilizes two crossbars with a hidden challenge passing from the first part to the second. We demonstrated that such feature could improve avalanche behavior and uniformity while maintaining other performance metrics close to ideal. Various PUF performance metrics have be analyzed. A uniformity of $47.28$\%, bit-aliasing of $47.48$\%, diffuseness of $49.86$\% and uniqueness of $49.85$\% are found. The PUF\rq{}s multiple column selection flexibility also offered a reliability of $98.67$\% under extreme process, voltage, temperature and sensing margin fluctuations. Additionally, we utilized a set of dummy ReRAMs to reduce {\it nr}PUF\rq{}s supply power SNR, although our read-out circuitry resulted in no meaningful relationship between power consumption and output bit generation of \lq\lq{}1\rq\rq{} or \lq\lq{}0\rq\rq{}. ReRAM devices in {\it nr}PUF are programmed in their HRS to (1) take advantage of highly spatially driven variations in HRS and (2) reduces power consumption. Crossbar aspects such as resistance-pattern dependent sneak current paths (parasitic current via neighboring cells) are also intrinsically contributing to {\it nr}PUF performance but their specific role in {\it nr}PUF operation is currently under investigation.


\end{document}